# Understanding Variations in Circularly Polarized Photoluminescence in Monolayer Transition Metal Dichalcogenides

*Kathleen M. McCreary,[1]\* Marc Currie,[1] Aubrey T. Hanbicki,[1] Berend T. Jonker[1]*

[1] *Naval Research Laboratory,* Washington DC 20375, USA

\* Author Information:

Correspondence and requests for materials should be addressed to K.M.M. (email: kathleen.mccreary@nrl.navy.mil)

Monolayer transition metal dichalcogenides are promising materials for valleytronic operations. They exhibit two inequivalent valleys in the Brillouin zone, and the valley populations can be directly controlled and determined using circularly polarized optical excitation and emission. The photoluminescence polarization reflects the ratio of the two valley populations. A wide range of values for the degree of circularly polarized emission, $P_{circ}$, has been reported for monolayer $WS_2$, although the reasons for the disparity are unclear. Here we optically populate one valley, and measure $P_{circ}$ to explore the valley population dynamics at room temperature in a large number of monolayer $WS_2$ samples synthesized via chemical vapor deposition. Under resonant excitation, $P_{circ}$ ranges from 2% to 32%, and we observe a pronounced inverse relationship between photoluminescence (PL) intensity and $P_{circ}$. High quality samples exhibiting strong PL and long exciton relaxation time exhibit a low degree of valley polarization, and vice versa. This behavior is also demonstrated in monolayer $WSe_2$ samples and transferred $WS_2$, indicating that this correlation may be more generally observed and account for the wide variations reported for $P_{circ}$. Time resolved PL provides insight into the role of radiative and non-radiative contributions to the observed polarization. Short non-radiative lifetimes result in a higher measured polarization by limiting opportunity for depolarizing scattering events.

Monolayer transition metal dichalcogenides such as $WS_2$ are direct gap semiconductors that have demonstrated high electronic mobility,[1] high optical responsivity,[2] and strong sensitivity to the surrounding environment, making them promising materials for future electronic and optoelectronic technologies, as well as chemical sensing.[3] Additionally, monolayer $WS_2$ has unique properties that make it an ideal candidate for valleytronic operations.[4,5] Analogous to spintronics, the goal of valleytronics is to utilize the valley degree of freedom in order to store information and control device operations.

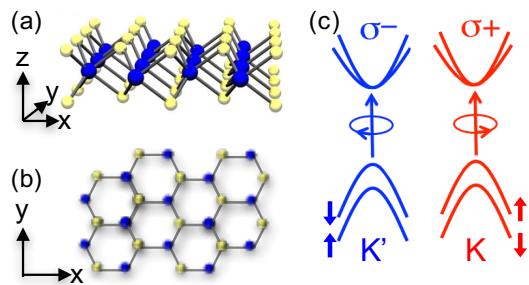

**Figure 1: Schematic of $WS_2$ atomic structure and valley dependent optical selection rules.** (a,b) The hexagonal lattice, as viewed normal to the surface, is composed by sandwiching a layer of tungsten (blue) with a top and bottom sulfur sheet (yellow). (c) Schematic diagram of the single-particle electronic band structure and valley dependent optical selection rules.

$WS_2$ is a layered structure with each layer consisting of a plane of tungsten atoms sandwiched by a top and bottom sulfur layer (Figure 1a). Monolayer $WS_2$ exhibits a hexagonal lattice structure, as viewed normal to the surface (Figure 1b), resulting in inequivalent K and K' valleys at the edges of the Brillouin zone. These energy degenerate valleys have a large valence band (VB) spin splitting of ~420 meV for $WS_2$,[6,7] and are distinguished by the spin orientation of the valence band maximum (VBM). Because of strong spin-orbit coupling and time reversal symmetry, the top of the valence band of $WS_2$ is spin up (spin down) in the K (K') valley (Figure 1c). This difference results in valley-dependent optical selection rules. Circularly polarized light with positive helicity (σ+) couples to excitonic transitions in the K valley, whereas negative helicity light (σ-) couples solely to the K' valley.[4,8–10]

These selection rules allow one to optically populate a specific valley with near-resonant excitation as well as probe the subsequent valley populations upon radiative decay. Optical excitation using circularly polarized light produces excitons (electron-hole pairs bound by the Coulomb interaction) in a single valley. Once created, Coulombic interactions,[11,12] phonon mediated scattering,[8,13] or coupling with atomic scale magnetic scatterers[5] can promote intervalley scattering between K and K', modifying valley populations. Radiative recombination of excitonic species is governed by the same valley-dependent optical selection rules. Therefore, measuring the circular polarization ($P_{circ}$) of the photoluminescence (PL) provides a direct way to monitor valley populations.

A high degree of valley polarization has been theoretically predicted for resonant excitation of monolayer transition metal dichalcogenides (TMDs),[4,5,10] but the expected high circular polarizations are rarely observed experimentally. In addition, the published values for $P_{circ}$ vary widely. For monolayer $MoS_2$ resonantly excited at 633 nm (1.96 eV), $P_{circ}$ ranges from 23% to 100% at cryogenic temperatures and 0% to 40% at room temperature.[4,9,14–16] While considerably less studied, $WS_2$ monolayers have exhibited similar variability, with $P_{circ}$ values of 12% and 40% reported at cryogenic temperatures and 0% and 10% at room temperature under near-resonant excitation of 594 nm (2.09 eV).[17,18] It is of fundamental importance to determine the origin of these variations to understand the basic mechanisms that govern exciton behavior.

We report here on the room temperature, circularly-polarized PL from monolayer $WS_2$ and $WSe_2$ on $SiO_2/Si$ substrates. Spatially resolved images as well as point spectra of polarized emission are acquired across numerous samples synthesized using chemical vapor deposition (CVD). Data are acquired from samples directly on the growth substrate and also following transfer to a second substrate. Substantial variation is observed in both $P_{circ}$ and PL intensity. For as-grown $WS_2$, $P_{circ}$ values range from 0% to 20% under 588 nm (2.11 eV) excitation, and the PL intensity varies over an order of magnitude. An inverse relationship between PL intensity and $P_{circ}$ is observed, with strong exciton luminescence observed in samples having a small degree of valley-polarized emission. Exciton decay dynamics are measured for representative samples using time-resolved PL, and the longest-lived decay component is found to range from 300 ps to ~1.5 ns. We show there is a strong correlation between exciton lifetime, $P_{circ}$, and PL intensity. Specifically, large valley-polarized emission ($P_{circ}$) is observed in samples exhibiting short exciton lifetimes. Our results suggest increased sample disorder and non-radiative intravalley relaxation (rather than intervalley relaxation) causes these shorter excitonic lifetimes.

**Results**

Monolayer tungsten disulfide is synthesized on $SiO_2/Si$ (275nm thickness of $SiO_2$) substrates using chemical vapor deposition (CVD), as reported in our previous work.[19,20] Growth on $SiO_2/Si$ results in equilateral triangle shapes, characteristic of crystalline growth. The triangular flakes are randomly oriented on the growth substrate and typically exhibit lateral dimensions on the order of several tens of μm (Figure 2a), allowing us to optically probe many discrete locations across a single flake. Topographical characterization with an atomic force microscope (AFM) demonstrates high uniformity across the $WS_2$ flakes.[20]

Polarization resolved PL spectroscopy measurements are performed under excitation from a 588 nm continuous wave (CW) laser source. Figures 2b,c present the polarization resolved PL spectra acquired at the four distinct points indicated in Figure 2a.

At each location, σ+ helicity is used to excite the sample, and σ+ and σ- emission intensities are successively measured. The degree of circular polarization ($P_{circ}$) is determined by the expression

$$P_{circ} = \frac{I(\sigma_+) - I(\sigma_-)}{I(\sigma_+) + I(\sigma_-)} \quad (1)$$

where $I(\sigma_{+/-})$ are the polarization resolved PL intensities. The PL spectra (Figure 2b) all display a single peak with maximum intensity at ~1.96 eV, indicating the room temperature emission characteristics are dominated by the neutral exciton.[20] The spectral line shape and emission energy are nearly identical across the four measured locations (Figure 2b). In contrast, the intensity varies considerably, ranging from 685 ct/sec (point 3) to 1610 ct/sec (point 4). The polarization values at each location (Figure 2c) are also notably different, with values ranging from 6.9% to 13.1%.

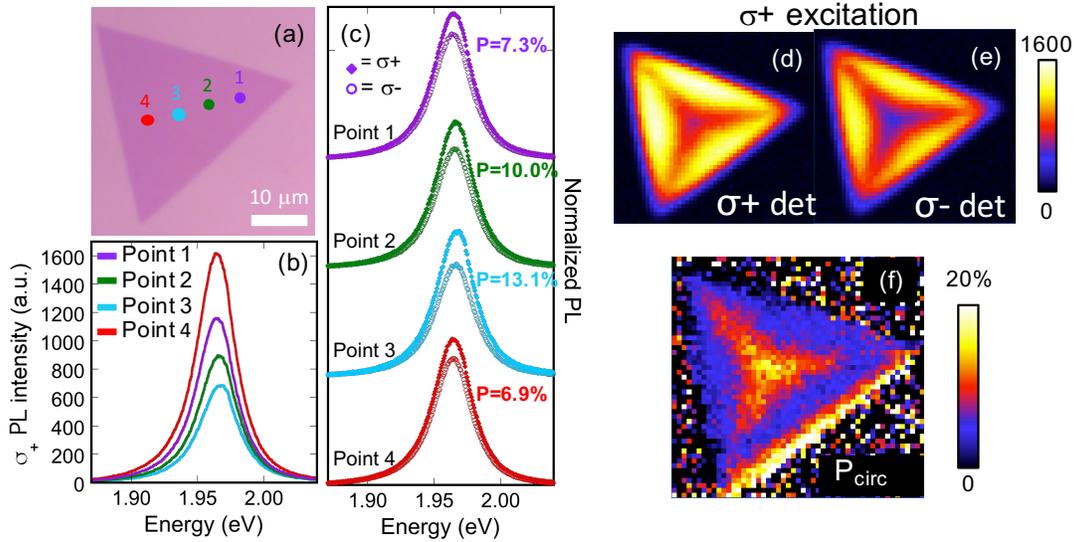

**Figure 2: Circularly polarized emission from as-grown monolayer WS$_2$ under σ+ 588 nm excitation.** (a) An optical micrograph of the investigated monolayer triangle. The colored circles represent the laser beam size and measurement locations. (b) The emission intensity and (c) σ+ (solid diamond) and σ− (open circle) PL components are measured at each of the four points. Data in (c) are normalized to the σ+ emission and offset for clarity, with $P_{circ}$ values listed to the right of each curve. Intensity maps are obtained across the entire monolayer for (d) σ+ polarized emission and (e) σ− polarized emission and used to determine (f) the degree of valley polarization, $P_{circ}$. A clear inverse relation between PL intensity and $P_{circ}$ is present across the monolayer flake.

A comparison of the PL intensity and the degree of valley polarization data at points 1-4 suggests an inverse relationship may exist – samples with the highest emission intensity exhibit the lowest polarization (point 4, red) and those with the lowest intensity have the highest polarization (point 3, blue). To further investigate this behavior, polarization-resolved PL mapping is acquired across the entire triangular flake. For map acquisition, the sample is excited using σ+ helicity and the σ+ component of PL emission is measured at each pixel as the laser rasters across the entire 46 x 43 μm region at 0.9 μm steps. The detection optics are then set for σ- analysis and the same area is rescanned.

The maximum peak intensity for σ+ (σ-) detection within the range of 1.9 eV to 2.1 eV is determined and plotted in Figure 2d (2e). Maps of the σ+ and σ- intensity show the same qualitative behavior, with low intensity present at the center of the triangle and radiating outward toward the three vertices. Similar intensity profiles have been reported in CVD synthesized materials for unpolarized excitation and detection[21-23] and is likely caused by variations in structural and chemical defects across the sample.[23] The degree of circular polarization across the sample is determined from equation (1) and displayed in Figure 2f. In direct contrast to the PL intensity, the highest degree of polarization is measured at the center of the triangle and decreases monotonically along the line between the center and each of the three vertices. The regions of lowest polarization correspond to those with the highest PL intensity. Polarization-resolved mapping is also acquired using σ- excitation rather than σ+, and we obtain the same result (Supplementary Information).

To determine whether the observed relationship between emission intensity and circular polarization applies in general, we performed similar measurements on $WS_2$ samples synthesized in seven different growth runs. The $WS_2$ samples exhibit a wide range of emission intensities, spanning over an order of magnitude, from 500 to 25,000 counts per second (Figure 3a). All samples shown in Figure 3a are synthesized on identical substrates, excited with the same laser power and wavelength, and measured at room temperature in air. We note that measurements using unpolarized excitation and detection exhibit similar intensity variation (not shown). The range of measured polarization values also expands, with $P_{circ}$ values ranging from 0.2% to 19.7% under σ+, 588nm laser-excitation ($\lambda_{exc}$). Figure 3a is a compilation of data from all of our samples and confirms a qualitatively inverse relationship between PL emission intensity and $P_{circ}$. The dashed line provides a guide to the eye.

Representative samples exhibiting low, moderate, and high $P_{circ}$ are selected for further investigation and are indicated as S1, S2, and S3 in Figure 3a. The inset of Figure 3a shows the polarization resolved spectra from samples S1 and S2. Sample S3 corresponds to the data collected in Figure 2 (point 3, blue line). Recent reports have demonstrated a strong sensitivity to laser excitation conditions, with polarization increasing as laser wavelength approaches resonance.[8,13] In fact, room temperature circular polarization as high as 35% has been reported in CVD synthesized monolayer $WS_2$ using 633 nm excitation.[24] We observe similarly high polarization values for S3, in which $P_{circ}$ increases from 13.1% for 588nm excitation to 32% for 633 nm excitation (Figure 3b). When exciting with 633nm light and measuring the exciton emission, an ultra-steep long pass edge filter (Semrock 633nm RazorEdge) is necessary to eliminate scattered laser light. The filter allows us to obtain PL spectra for resonant excitation, but consequently truncates the spectra above 1.955eV, as evident in Figure 3b. Samples S1 and S2 likewise exhibit an enhancement, from 0.2% to 2% and from 7.6% to 23% respectively, when utilizing resonant excitation.

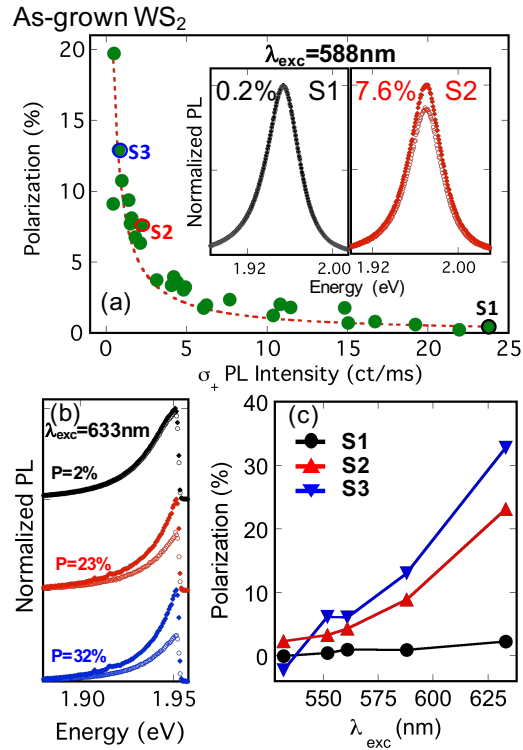

**Figure 3: Valley polarization across multiple as-grown WS$_2$ samples.** The circularly polarized emission is measured at 28 locations from samples produced in 7 separate CVD synthesis runs. The P$_{circ}$ vs. PL emission intensity is presented in (a) with the red dashed line provided as a guide to the eye. The insets display polarization resolved emission from sample S1 (black) and S2 (red). (b) Select locations S1, S2, S3 indicated in (a) are further investigated with resonant excitation at 633 nm. The PL spectra are truncated at high energy due to the filter used to eliminate scattered laser light. Solid (open) points represent PL analyzed as σ+(σ-). (c) P$_{circ}$ as a function of laser excitation energy for the same positions S1,S2,S3.

A more complete picture of the excitation energy dependence is formed by measuring the response to 561 nm, 552 nm, and 532 nm laser excitation and incorporating these data with the above presented data collected with 633 nm and 588 nm excitation (Figure 3c). P$_{circ}$ decreases as the separation between WS$_2$ emission energy and laser excitation increases. This observed behavior is consistent with various valley dephasing mechanisms reported in other work including direct intervalley scattering[8,13,25] and electron-hole exchange interactions.[11,26] While the specific emission vs. polarization values are distinct for each sample, the overall excitation dependent trend is the same for all three samples.

Strain is known to significantly impact the properties of monolayer TMD samples. To determine whether the pronounced inverse relationship between $P_{circ}$ and PL intensity is related to strain that exists in the as-grown samples on the growth substrate, we transfer these samples to a second, identical Si/SiO$_2$ substrate as described in reference [20]. Recent works have found that fundamental characteristics of as-grown and transferred samples can be quite different.[20,27] In particular, the energy of PL emission changes. While as-grown WS$_2$ (as-WS$_2$) emits at ~1.96 eV, the PL emission peak shifts above 2 eV after these samples are transferred to a second identical Si/SiO$_2$ substrate (x-WS$_2$). This is clearly seen in the PL spectra shown in the inset of Figure 4. This disparity has been attributed to differences in tensile strain which is known to reduce the band-gap of monolayer TMDs.[28,29] Such strain is present in as-WS$_2$ but is removed during the transfer process. To the best of our knowledge, a comparison of valley polarization between as-WS$_2$ and x-WS$_2$ has not yet been presented.

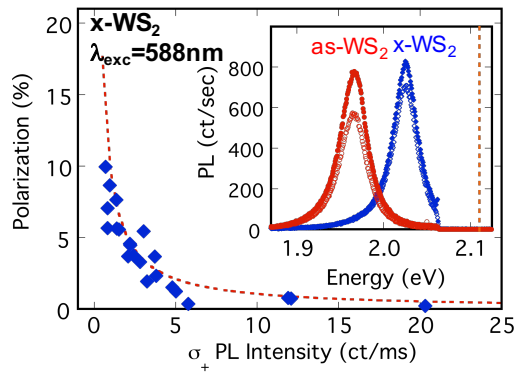

**Figure 4: Polarization of transferred CVD WS$_2$.** The inverse relation between polarization and PL intensity remains in transferred films. Samples are investigated using identical conditions as Fig. 3a and the red dashed line indicates the behavior observed in as-grown WS$_2$. The inset shows the polarization-resolved PL for representative as-grown and transferred WS$_2$ samples.

The PL intensity and circular polarization are acquired for 22 x-WS$_2$ samples and presented in Figure 4. The excitation conditions are identical to those used for the as-WS$_2$ data presented in Figure 3a, and the guide-line from Figure 3a (red-dashed) is replicated in Figure 4 for comparison. Notably, the un-strained x-WS$_2$ samples exhibit the same qualitatively inverse relationship between P$_{circ}$ and PL intensity observed for the as-WS$_2$ samples under tensile strain, ruling out variations in strain as the source of the observed inverse relationship.

A similar inverse relationship between P$_{circ}$ and PL intensity is observed in monolayer WSe$_2$ samples, indicating the behavior and mechanisms are relevant to other TMD materials. Samples are synthesized on SiO$_2$/Si substrates using similar procedures employed for WS$_2$ growth, but with an elemental selenium precursor and hydrogen flow throughout the entire process. Strong PL emission combined with Raman spectroscopy confirm the monolayer nature of investigated samples (supplementary information). The circular polarization and PL intensity of as-grown monolayer WSe$_2$ is acquired as described above using $\lambda_{exc}$=685 nm (1.81 eV) across multiple samples and summarized in Figure 5. Sample S4, indicated in the figure, exhibits the lowest measured intensity and highest polarization (P$_{circ}$ = 16%) among the measured WSe$_2$ monolayers, whereas S5 demonstrates the highest PL intensity and lowest P$_{circ}$ = 2%. The corresponding spectra are shown in the insets. The inverse relationship between PL intensity and P$_{circ}$ is clearly evident in this selenium-based TMD as well.

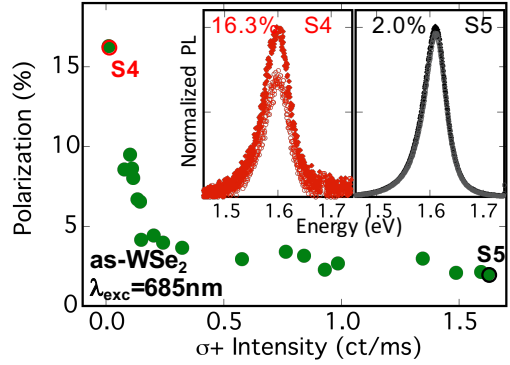

**Figure 5: Valley polarization in monolayer WSe$_2$ under σ+ 685 nm excitation.** The relationship between P$_{circ}$ and σ+ intensity exhibits an inverse relationship. The inset shows PL spectra from monolayer samples having low and high intensity emission.

**Discussion**

While intervalley scattering may explain the general excitation-energy dependence of P$_{circ}$ shown in Figure 3c, it cannot account for the large variation in P$_{circ}$ observed under identical excitation conditions or the large variations in PL intensity (Figure 3a). Using a rate equation model,[4,17] for steady state conditions achieved with CW laser excitation, the circular polarization, *P$_{circ}$*, is given by

$$P_{circ} = \frac{P_0}{1+2\frac{\tau_e}{\tau_s}} \qquad (2)$$

where *P$_0$* is the initial polarization, *τ$_e$* is the exciton relaxation time and *τ$_s$* is the valley relaxation time. Thus, P$_{circ}$ will increase with either a decrease in exciton relaxation time or an increase in the valley relaxation time. Because all as-WS$_2$ samples investigated in Figure 3a are synthesized on identical substrates, excited with the same laser power and wavelength, and exposed to the same environment, we assume that the optically generated exciton density and *P$_o$* are the same. The exciton relaxation time depends upon both radiative and non-radiative recombination through the relationship

$$\frac{1}{\tau_e} = \frac{1}{\tau_r} + \frac{1}{\tau_{nr}} \qquad (3)$$

where $\tau_r$ is the radiative lifetime and $\tau_{nr}$ the non-radiative lifetime. In monolayer transition metal dichalcogenides at room temperature, non-radiative lifetimes can be orders of magnitude shorter than radiative recombination,[30,31] thereby reducing the overall exciton relaxation time. The increase in $P_{circ}$ and concurrent reduction in PL intensity as evident in Figures 3 and 4 are therefore consistent with a decreasing exciton relaxation time resulting from a decreasing non-radiative lifetime. These observations suggest variations in valley polarization are dominated by differences in $\tau_e$ from sample-to-sample, and that additional disorder in these system leads to higher polarization due to shorter $\tau_{nr}$. A similar phenomenon has recently been shown in a III-V quantum well system.[32]

To support this hypothesis, time resolved photoluminescence (TRPL) is utilized to determine the temporal evolution of the PL and the corresponding radiative lifetimes in as-grown $WS_2$ samples S1 and S2, which exhibit significantly different PL intensities and polarizations (Figure 3a). The TRPL measurements are performed using pulsed, linearly-polarized light of 532 nm, creating excitons equally in both K and K' valleys. Subsequent luminescence from both valleys is collected and directed to a time-resolved point detector. Just as samples S1 and S2 displayed markedly different PL intensity vs. polarization behavior (Figure 3), the time-resolved emission spectra of these samples also exhibit noticeably different exciton decay dynamics (Figure 6). The sample that exhibits higher PL intensity (i.e. S1) has longer exciton lifetime, while the one with weaker PL (S2) displays shorter lifetime.

The luminescence data are fit with a multi-exponential decay of the form

$$I(t) = \sum_{i=1}^{N} a_i e^{-t/\tau_i} \qquad (4)$$

convolved with the instrument response function (Figure 6, green curve), where $\tau_i$ and $a_i$ correspond to the lifetime and the fractional contribution of each component to the decay, respectively.[31] The temporal evolution of the PL intensity of sample S2 is well fit by a two-term exponential composed of a fast initial decay, $\tau_1 \sim 30$ ps, followed by a moderate $\tau_2$ $\sim 300$ ps decay. The lifetime associated with sample S1 has similar components, but in addition has a much longer-lived component. This is modeled by expanding to a three-term exponential fit. The resulting fit yields $\tau_1 \sim 50$, $\tau_2 \sim 300$ ps, and $\tau_3 \sim 1500$ ps. These components are summarized in the table insert in Figure 6.

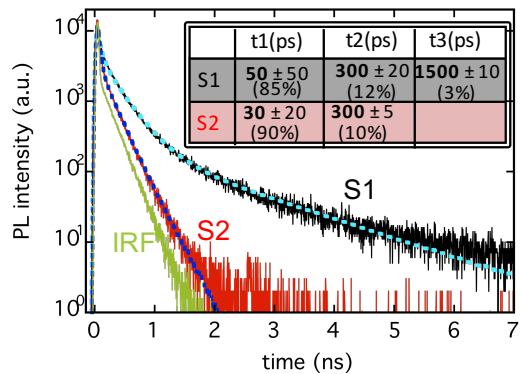

**Figure 6: Time resolved PL emission.** Samples S1 and S2 are excited with linearly polarized 532 nm light and time resolved PL emission is presented. Data are well fit by multi-component exponentials (blue dashed lines) convolved with the instrument response function (IRF). The lifetime and fractional amplitude of each fit component are displayed in the inset.

These decay components can be attributed to a variety of sources. At room temperature, various non-radiative mechanisms are expected to compete with radiative recombination. Potential avenues for non-radiative decay arise from disorder (which can originate from grain boundaries, traps, localized states and recombination centers) and energy equilibration processes (such as Auger recombination, hopping, and dipole-dipole interactions)[33–36]. Such non-radiative mechanisms are likely the source of the short $\sim 30$ps and moderate $\sim 300$ps decay dynamics.[31,34,37] The long-lived $\sim 1.5$ns decay, observed only in sample S1, is attributed to radiative recombination $\tau_r$ and coincides well with the theoretically predicted values of several ns.[30,33,38] Additionally, the fractional amplitude corresponding to radiative recombination in S1 is found to be 0.03 (see insert Figure 6), consistent with recent quantum yield measurements on the order of a few percent in monolayer $WS_2$.[35]  In comparing samples S1 and S2, it is clear that the sample exhibiting shorter exciton recombination time and higher disorder (S2) has higher circular polarization, as expected. Short non-radiative lifetimes result in a higher measured PL polarization by limiting opportunity for depolarizing scattering events.

**Summary and Conclusions**

In conclusion, we have investigated the room-temperature circularly-polarized PL in several TMD systems, including as-grown $WS_2$, as-grown $WSe_2$, and transferred $WS_2$ monolayers. In each system, a wide range of $P_{circ}$ and PL intensity values are observed from sample to sample.  We show that there is a pronounced inverse correlation between $P_{circ}$ and PL intensity, which we attribute to sample dependent variations in the exciton radiative and non-radiative lifetime components.  High values of $P_{circ}$ are correlated with short non-radiative lifetimes which we associate with a higher degree of sample disorder arising from extrinsic effects such as defects and trap states.  Samples with strong PL intensity and long radiative lifetimes exhibit low valley-polarization. These findings clarify the disparities among previously reported values and suggest a means to engineer valley polarization via controlled introduction of defects and non-radiative recombination sites.

**Methods:**

   **Monolayer TMD synthesis.** Chemical vapor deposition (CVD) synthesis of monolayer $WS_2$ is performed at ambient pressure in a 2-inch diameter quartz tube furnace on $SiO_2$/Si substrates (275 nm thickness of $SiO_2$). Prior to use, all growth substrates undergo a standard cleaning procedure consisting of (i) ultrasonic cleaning in acetone, (ii) ultrasonic cleaning in isopropyl alcohol, (iii) submersion in Piranha etch (3:1 mixture of $H_2SO_4$/$H_2O_2$) for approximately 2 hours, and (iv) thorough rinsing in deionized water. A quartz boat containing ~1 g of $WO_3$ powder was positioned at the center of the furnace. Two $SiO_2$/Si wafers are positioned face-down, directly above the oxide precursor. The upstream wafer contains perylene-3,4,9,10- tetracarboxylic acid tetrapotassium salt (PTAS) seeding molecules,[39] while the downstream substrate is untreated. The hexagonal PTAS molecules are carried downstream to the untreated substrate and promote lateral growth of $WS_2$. A separate quartz boat containing sulfur powder is placed upstream, outside the furnace-heating zone. Pure argon (65 sccm) is used as the furnace heats to the target temperature. Upon reaching the target temperature of 825 °C, 10 sccm $H_2$ is added to the Ar flow and maintained throughout the 10 min soak and subsequent cooling. Synthesis of monolayer $WSe_2$ is performed in a separate 2-inch diameter tube furnace. The growth substrates, cleaning procedure, and seeding molecules used in $WSe_2$ growth are identical to those utilized for $WS_2$ synthesis. The $WO_3$ is heated to 825°C under continuous flow of 65 sccm Ar and 10 sccm $H_2$, with selenium powder positioned upstream.

**$WS_2$ transfer process.** The CVD-grown $WS_2$ samples on $SiO_2$ are transferred onto a clean $SiO_2$/Si substrate using a PMMA-assisted technique. A thin layer of polymethyl methacrylate (PMMA) is spun onto the surface of the entire growth substrate then submerged in buffered oxide etchant. After several hours, the oxide layer is removed, freeing the $WS_2$/PMMA film from the growth substrate. The sample is subsequently transferred to $H_2O$ to rinse chemical etchants, where the fresh Si/$SiO_2$ is used to lift the film out of the water. A 2000 rpm spin and 150 °C bake improve the uniformity and adhesion to the substrate, after which the PMMA is dissolved in acetone.

**Photoluminescence Spectroscopy**. Photoluminescence spectra are acquired at room temperature in ambient conditions using a commercial Horiba LabRam confocal spectrometer. A 50× objective is used to focus the laser beam to a spot of ~2 μm diameter. Polarization resolved PL spectroscopy measurements are performed under excitation from a continuous wave (CW) laser source. A quarter wave plate (Thorlabs superachromatic) is used to circularly polarize the laser excitation. The resulting photoluminescence is collected and directed through the same quarter wave plate and a subsequent rotatable linear polarizer to analyze the circularly polarized emission components. Beam steering mirrors control the laser position in the xy sample plane and enable both single spot and scanned area acquisition.

**Time Resolved Photoluminescence** Tunable wavelength optical pulses were produced by optical parametric amplification using a Ti:sapphire laser. The output pulses were filtered to ~1-nm bandwidth using a scanning monochromator, reflected off a long-pass optical filter and were focused onto the sample using a 50x microscope objective. The luminescence was collected using the same objective and passed through the long-pass filter and was imaged in a confocal setup to a spectrometer (Horiba iHR-320). The spectrometer had two exit ports: one for an imaging camera, and another for a time-resolved single-photon counter (Micro Photon Devices PDM). The time-resolved instrument response was measured using scattered light from the sub-picosecond excitation beam, and was identical to that of the scattered light when the pulsed laser was tuned to 620nm. Time resolved data were recorded and post-processed for deconvolution of the instrument response to achieve the sample response.

**Acknowledgements**
The authors acknowledge helpful discussions with Paul Cunningham and George Kioseoglou. Core programs at NRL and the NRL Nanoscience Institute supported this work. This work was supported in part by the Air Force Office of Scientific Research under contract number AOARD 14IOA018-134141.

**Supporting Information Available:** Circularly polarized emission of as-grown $WS_2$ under $\sigma$- excitation. Raman characterization of CVD synthesized $WSe_2$. This material is available free of charge *via* the Internet at http://pubs.acs.org.